\documentclass[12pt]{article}
\usepackage{amsmath}
\usepackage{enumerate}
\usepackage{amsfonts}

\usepackage{amsmath}
\usepackage{mathrsfs}
\usepackage{enumerate}
\usepackage{amssymb}
\usepackage{amsmath}

\begin{document}
\def\be{\begin{equation}}
\def\bea{\begin{eqnarray}}
\def\ee{\end{equation}}
\def\eea{\end{eqnarray}}
\def\d{\partial}
\def\eps{\varepsilon}
\def\la{\lambda}
\def\b{\bigskip}
\def\tgt{}  
\def\nn{\nonumber \\}
\def\t{\tilde}
\def\varkappa{\kappa}

\def\nn{\nonumber\\ }
\def\eps{\epsilon}

\def\RR{\mathscr{R}}
\def \r{\mathscr{R}}
\def\TT{t}
\def\d{\partial}
\def\la{\lambda}

\makeatletter
\def\blfootnote{\xdef\@thefnmark{}\@footnotetext}  
\makeatother

\begin{center}
{\LARGE Analytical Structure of the Generalized $\lambda-$deformation}
\\
\vspace{18mm}
{\bf Oleg Lunin and Wukongjiaozi Tian}
\vspace{14mm}

Department of Physics,\\ University at Albany (SUNY),\\ Albany, NY 12222, USA\\ 

\vskip 10 mm

\blfootnote{olunin@albany.edu,~jtian2@albany.edu}

\end{center}

\begin{abstract}

We explore the analytical structure of the generalized $\lambda-$deformation of AdS$_p\times$S$^p$ spaces and construct new integrable backgrounds which depend on $(p+1)$ continuous parameters.

\b

\end{abstract}

\newpage

\newpage
\section{Introduction}
The last two decades witnessed a remarkable progress in understanding of integrable string theories. Following its original discovery in isolated examples of AdS$_p\times$S$^q$ \cite{AdS5int,AdS3int,AdS2int}, integrability has been extended to continuous families of string theories known as beta--, eta-- and lambda--deformations \cite{beta,DelducEta,eta,SftsGr,MoreLmb,DSTLmbAdS5}. Interestingly, the eta-- and lambda-- families are nicely unified in the framework of so--called generalized lambda deformation \cite{GenLambda}, which will be the subject of this article.

The original  lambda--deformation \cite{SftsGr,MoreLmb,DSTLmbAdS5} provides a one--parametric interpolation between two well--known solvable CFTs, the Wess--Zumino--Witten and the Principal Chiral models. The generalized version introduces additional parameters by modifying this family using the solution of the Yang-Baxter equation, the main ingredient of the eta--deformation\footnote{See \cite{DelducEta,Cherednyk,BosonicYangBaxter,Qdeform,Yoshida} for the detailed discussion of connections between the Yang--Baxter equation and integrable deformations.}. In this article we will refer to the generalized lambda deformation using a shorthand label $\la$YB, which stresses two elements of the construction: the interpolation and the Yang--Baxter equation. The freedom in choosing a solution of the Yang--Baxter equation suggests that the generalized lambda--deformation ($\la$YB) should have many free parameters, whose number should grow with the size of the group. However, in the few explicit examples constructed in \cite{GenLambda,ChLGL} most of these parameters could have been absorbed in coordinate redefinitions, and this observation raises questions about usefulness of the generalized lambda deformation in comparison to its standard version. In this paper we construct the most general $\la$YB deformations of the $N$--dimensional spheres and demonstrate that such integrable systems are governed by $(N+1)$ continuous parameters. We will also show that the reduction of the number of parameters observed in  \cite{ChLGL} is an accident associated with low dimensionality of $S^2$ and $S^3$.

The $\la$YB deformations are parameterized by solutions of the classical Yang--Baxter equation (CYBE), 
which has been extensively studied in the mathematical literature \cite{BelavDrinf,CYB}. In this article we will apply the insights from such investigations to the problem of constructing the integrable $\la$YB deformations of the AdS$_p\times$S$^q$. The existing constructions \cite{GenLambda,ChLGL} used only the canonical solution of the CYBE and its simplest embedding into the coset, which led to a small number of deformation parameters. In this article we will extend these results to the {\it most general} $\la$YB deformations and demonstrate that the canonical solution of CYBE is indeed the only R--matrix that leads to an integrable system after the deformation\footnote{This statement has been proven perturbatively in \cite{ChLGL}, but here we will present a different argument which is not based on expansion in a small parameter.}, but different embeddings of this solution into the coset lead to $(N+1)$ deformation parameters.

\bigskip

This paper has the following organization. In section \ref{SectReview} we review the procedure for finding the generalized $\lambda-$deformation introduced in \cite{GenLambda}. This construction is based on solutions of the Yang--Baxter equations which satisfy an additional constraint, and finding the most general R--matrix with such properties is the main goal of this article. Specifically, we focus on deformations of spheres $S^N=SO(N+1)/SO(N)$ and AdS geometries. To ensure integrability of the deformed coset space, the R--matrix must satisfy the Yang--Baxter equation on the group $G$ and a projection associated with the coset. In section \ref{SecRealR} we construct the most general {\it real} 
R--matrix satisfying a Yang--Baxter equation for any {real} Lie group, such as $SO(N+1)$. In section \ref{SectMainDeform} we add the constraint associated with the coset to construct the most general integrable $\la$YB deformation of $S^N=SO(N+1)/SO(N)$. An extension to AdS spaces is accomplished by a simple analytic continuation. Some technical details are presented in the appendices.

\section{Review of the generalized $\lambda$-deformation}
\label{SectReview}
\renewcommand{\theequation}{2.\arabic{equation}}
\setcounter{equation}{0}

Lambda deformations of the Principal Chiral Models (PCM) were introduced in \cite{SftsGr} and further studied in \cite{MoreLmb,DSTLmbAdS5}. Application of such deformation to any PCM leads to a one--parameter family of integrable conformal field theories. More general families, which will be the main subject of this paper, were introduced in \cite{GenLambda}, and we begin with reviewing this construction following section 5 of \cite{GenLambda}.

\bigskip

The $\lambda$ deformation  interpolates between Conformal Field Theories described by a Wess--Zumino--Witten Model (WZW) \cite{WZW},
\bea\label{gWZW}
S_{WZW,k}(g)=\frac{k}{4\pi}\int_\Sigma d^2\sigma R_+^a R_-^a-\frac{k}{24\pi}\int_B f_{abc} R^a\wedge R^b\wedge R^c,\quad \d B=\Sigma,
\quad g\in G\,,
\eea
and a generalized Principal Chiral Model (PCM) \cite{PCM},
\bea\label{genPCM}
S_{gPCM}(\hat{g})=\frac{k}{2\pi}\int d^2\sigma {E}_{ab} R^a_+(\hat{g}) R^b_-(\hat{g}),\qquad
\hat{g}\in G\,.
\eea
Here $g$ and $\hat g$ are elements of some Lie group $G$ with generators $T_a$ and structure constants $f_{abc}$, $k$ is the level of the WZW model, and
$R_\pm$ are the right-invariant Maurer-Cartan forms,
\bea
R^a_\pm=-i \mbox{Tr}[T^a \d_\pm g  g^{-1}]\,.
\eea
To construct the $\lambda$ deformation one adds the actions (\ref{gWZW}) and (\ref{genPCM}),
and gauges away half of the degrees of freedom in the resulting 
sum\footnote{See \cite{GenLambda} for more details.}.
Parameters ${E}_{ab}$ in (\ref{genPCM}) represent a constant matrix, whose form will be restricted by the requirements of conformal invariance and integrability. The gauging procedure in the sum of (\ref{gWZW}) and (\ref{genPCM}) leads to the action \cite{GenLambda}
\bea\label{GenLambdaAction}
S_{k,\lambda}(g)=S_{WZW,k}(g)+\frac{k}{2\pi}\int d^2\sigma L_+^a[\hat{\lambda}^{-1}-D]^{-1}R_-^b,
\eea
where\footnote{Following \cite{GenLambda}, we denote the {\it matrix} appearing in (\ref{GenLambdaAction}), (\ref{lambdaMtrDef}) by $\hat{\lambda}$ to distinguish it from the {\it scalar} deformation parameter $\la$.}  
\bea\label{lambdaMtrDef}
\hat{\lambda}^{-1}={E}+ I,
\quad D_{ab}=\mbox{Tr}[T_a g T_b g^{-1}],\quad
L_\pm^a
=i\mbox{Tr}[T_ag^{-1}\d_\pm g],\quad R_\mu^a=D_{ab}L_\mu^b.
\eea
Relations (\ref{GenLambdaAction}) and (\ref{lambdaMtrDef}) can be used to recover the metric and the Kalb--Ramond field, and the dilaton of the deformed model is given by
\bea\label{dilaton}
e^{-2\Phi}=\mbox{det}[\hat{\lambda}^{-1}-D].
\eea
Application of this prescription to the standard PCM,
\bea\label{StandLambda}
{E}_{ab}=\frac{\kappa^2}{k}\delta_{ab},\quad \hat{\lambda}^{-1}=\frac{k+\kappa^2}{k}I,
\eea
leads to a one-parameter $\lambda$--deformation, and integrability of the corresponding conformal field theory (\ref{GenLambdaAction}) was demonstrated in \cite{SftsGr}. It is clear that the sigma model \eqref{GenLambdaAction} would not be integrable for a generic matrix ${E}$, but a large class of integrable models extending (\ref{StandLambda}) was found in \cite{GenLambda}. 

\bigskip

Specifically, the authors of \cite{GenLambda} demonstrated that the model (\ref{GenLambdaAction}) with 
\bea\label{EmatrYB}
{E}_{YB}=\frac{1}{{t}}\left( I-{\eta}\RR \right)^{-1}
\eea
was integrable, as long as the matrix $\RR$ satisfied the modified classical Yang-Baxter (mCYB) equation\footnote{The constant matrix $\RR$ satisfying the Yang-Baxter equation is called the Yang-Baxter operator or the R--matrix. In this paper we use both names.}
\bea\label{modYBeq}\label{myb}
[\RR A,{\RR} B]-\RR([\RR A,B]+[A,\RR B])=-c^2 [A,B],\quad A,B\in \mathfrak{g},\quad c\in\mathbb{C}.
\eea
The generalized Principal Chiral Model (PCM) (\ref{genPCM}) with matrix $E$ given by (\ref{EmatrYB}) is known as $\eta$--deformation, its integrability was demonstrated in \cite{BosonicYangBaxter}, and various aspects of the resulting geometries have been discussed in 
\cite{DelducEta,eta,Yoshida,xxx}\footnote{Recently an intriguing relation between the $\eta$--deformation and noncommutativity has been uncovered in \cite{SWmap}.}. The construction introduced in \cite{GenLambda} added an extra tunable parameter $\la$ to these families.

Article \cite{GenLambda} also extended the notion of the generalized $\lambda$-deformation to cosets $G/F$ by defining 
\bea\label{EmtrCosetA}
{E}={E}_H\oplus{E}_{G/F},\quad {E}_F=0,\quad {E}_{G/F}=\frac{1}{{t}}(I-{\eta}\RR)^{-1},\quad \mathfrak{g}=\mathfrak{f}+\mathfrak{l},
\eea
and demonstrating that integrability of \eqref{GenLambdaAction} for any R--matrix that satisfies equation (\ref{modYBeq}) and the 
constraint\footnote{This constraint is multiplied by $\eta$, but since we are interested in the deformed theory, $\eta\ne 0$.}:
\bea\label{YBcosetConstr}\label{constraint}
[X,Y]_\RR|_\mathfrak{f}=([\RR X,Y]+[X,\RR Y])|_\mathfrak{f}=0,\quad X,Y\in\mathfrak{l}.
\eea
In the rest of this article we will study some analytical properties of the construction (\ref{EmatrYB})--(\ref{YBcosetConstr}) for $G/F=SO(N+1)/SO(N)$.
Such cosets naturally arise in string theory as $N$--dimensional spheres, and extension to AdS spaces will be briefly discussed in the end of section \ref{SectMainDeform}.

\section{R--matrices for real algebras}
\label{SecRealR}
\renewcommand{\theequation}{3.\arabic{equation}}
\setcounter{equation}{0}

Before focusing on specific cosets describing spheres, in this section we will study some 
general properties of the R--matrices satisfying equation (\ref{modYBeq}) on real algebras.
Relation (\ref{modYBeq}) is known as the modified Classical Yang--Baxter Equation (mCYBE) \cite{CYB}, and we are interested in its antisymmetric solutions:
\bea\label{YBanti}
{\RR}_{ab}=-{\RR}_{ba}.
\eea
The standard CYBE is obtained by setting $c=0$ in (\ref{modYBeq}). The mCYBE and CYBE have been subjects of extensive studies in mathematical literature \cite{BelavDrinf,CYB}, and a nice pedagogical review can be found in \cite{Guide}. In particular, it can be shown that any antisymmetric solution of the mCYBE (\ref{myb}) can be written as 
\bea
{\RR}=r-{\TT},
\eea
where $r$ satisfies the homogeneous CYBE, and ${\TT}$ is the Casimir element 
\bea\label{Casimir}
t=ic \sum T_A\otimes T_A\,.
\eea
All solutions of the CYBE were classified by Belavin and Drinfeld, and for any algebra they can be constructed using the following algorithm \cite{BelavDrinf,Guide}:
\label{BelDrinClassPage}
\begin{enumerate}
\item{Introduce a  Cartan-Weyl basis $\{H_i,E_\alpha,E_{-\alpha}\}$ and fix a set $\Pi$ of positive simple roots.}
\item{Define the admissible triples $(\Pi_1,\Pi_0,\tau)$ satisfying three conditions:}
\begin{enumerate}
	\item $\Pi_1,\Pi_0 \subset \Pi$.
	\item $\tau:\Pi_1 \rightarrow \Pi_0$ is a one-to-one product--preserving map: $\langle\tau(\alpha),\tau(\beta)\rangle=\langle\alpha,\beta\rangle$.
	\item for every $\tau\in \Pi_1$, there exists $m\in \mathbb{N}$ such that 
	$$\alpha,\tau(\alpha),\dots,\tau^{m-1}(\alpha)\in\Pi_1,\quad \mbox{but}\quad \tau^m(\alpha)\in \Pi_0$$
\end{enumerate}
\item{Define the partial order as $\beta>\alpha$ if $\tau^m(\alpha)=\beta$.}
\item{Any triple defined at step 2 leads to a solution of the CYBE 
\bea\label{CYBrSln}
r={\tilde r}^0+i\sum_{\alpha}E_{-\alpha}\otimes E_{\alpha}+i\sum_{\alpha,\beta,\alpha<\beta}
E_{-\alpha}\wedge E_{\beta}\,,
\eea
where the sum is extended over positive roots, and
${\tilde r}_0=\sum C_{ij}H_i \wedge H_j$ satisfies  a system of equations 
\bea
\Big[\tau(\alpha)\otimes 1+1\otimes \alpha\Big]{\tilde r}^0=0.
\eea
}
\end{enumerate}
While the normalization of the Casimir element (\ref{Casimir}) is fixed by choosing $c=-i$ in (\ref{myb}), 
the matrix $r$ in (\ref{CYBrSln}) satisfies a homogeneous equation, so it can be multiplied by any constant.
Only one such constant makes matrix ${\RR}=r-{\TT}$ antisymmetric:
\bea\label{mCYBgen}
\r={r}^0+i\sum_{\alpha}E_{\alpha}\wedge E_{-\alpha}-2i\sum_{\alpha,\beta,\alpha<\beta}E_{-\alpha}\wedge E_{\beta}\,.
\eea
The $\RR-$operator is obtained by tracing out the second entry in an orthonormal basis:
\bea\label{RmatrReadOff}
\RR X=\text{Tr}_2[\r (1\otimes X)]\equiv a_i \text{Tr}(b_i X)-b_i\text{Tr}(a_i X),~~r=a_i \wedge b_i.
\eea 
Every algebra admits a triple with $\Pi_0=\Pi_1=\emptyset$, which leads to the ``canonical solution'',
\bea\label{mCYBcan}
\r=r^0+i\sum_{\alpha}E_{\alpha}\wedge E_{-\alpha}\,,
\eea
where $r_0$ acts on the Cartan subspace. This solution and its properties will be discussed in the next section. Let us now demonstrate that any other solution leads to a complex R--matrix, and thus it cannot be used for deforming real cosets. As we will show in a moment, for real algebras, 
${\bar E}_\alpha=E_{-\alpha}$, then equation (\ref{mCYBgen}) gives
\bea
\r-{\bar \r}=r^0-{\bar r}^0-2i\sum_{\alpha,\beta,\alpha<\beta}[E_{-\alpha}\wedge E_{\beta}-E_{-\beta}\wedge E_{\alpha}]\,,
\eea
and the right--hand side must vanish for real solutions of the CYBE. Since the products $H_i \wedge H_j$ and $E_{\alpha}\wedge E_{\beta}$ are linearly independent, the sum over $\alpha<\beta$ should disappear in the last expression, and this happens only for the canonical solution. To complete this argument, we will now demonstrate that 
${\bar E}_\alpha=E_{-\alpha}$.

\bigskip

For any real simple algebra, we begin with choosing the (real) Cartan generators $H_i$. Then the roots $\alpha$ and ladder operators $E_\alpha$ are defined by
\bea\label{HEcommut}
[H_i,E_\alpha]=\alpha_i E_\alpha\,.
\eea 
The left--hand side of the last expression encodes the action of $H_i$ in the adjoint representation, and 
for a real algebra, the matrix $(H^{adj}_i)_{ab}=f_{iab}$ is real and antisymmetric. This implies that all eigenvalues of this matrix, i.e., all roots $\alpha_i$ of any real algebra, are imaginary. Thus the generator $E_\alpha$ must be complex, and it can be written as 
\bea
E_\alpha=E_\alpha^R+iE_\alpha^I.
\eea 
Then relation (\ref{HEcommut}) implies
\bea\label{EalRuleOut}
[H_i,E^R_\alpha]=i\alpha_i E^I_\alpha,\quad [H_i,E^I_\alpha]=-i\alpha_i E^R_\alpha,
\eea
and general properties of Lie algebras ensure that $(\la E^R_\alpha,\la E^I_\alpha)$ are the only real elements of the algebra satisfying the relations above\footnote{If there were another pair $({\tilde E}^R_\alpha, {\tilde E}^I_\alpha)$, then ${\tilde E}^R_\alpha+i{\tilde E}^I_\alpha$ would give a second ladder operator with root $\alpha$, which is not allowed.}. On the other hand, writing the real and imaginary parts of $E_{-\alpha}$ as
\bea
E_{-\alpha}={\tilde E}_\alpha^R-i{\tilde E}_\alpha^I\,,
\eea 
we observe that $({\tilde E}^R_\alpha, {\tilde E}^I_\alpha)$ satisfy (\ref{EalRuleOut}). This implies that $E_{-\alpha}=\la{\bar E}_\alpha$, where $\la$ is a {\it real} constant. By an appropriate rescaling of $E_{-\alpha}$ (which does not modify the form of commutation relations in the algebra), we can set $\la=1$. In other words, for any simple real algebra, $E_{-\alpha}={\bar E}_\alpha$, and an extension to the semi--simple case is straightforward.

\bigskip

To summarize, in this section we used the general properties of the mCYBE to show that  for a real group, the only real solution of (\ref{modYBeq}) is the canonical one, (\ref{mCYBcan}). The properties of integrable deformations generated by this solution will be investigated in the remaining part of the paper. 

\section{Deformations of the $SO(N+1)/SO(N)$ coset}
\renewcommand{\theequation}{4.\arabic{equation}}
\setcounter{equation}{0}
\label{SectMainDeform}

Although the general construction reviewed in section \ref{SectReview} can be applied to any coset, the most interesting integrable string theories are associated with AdS$_p\times$S$^p$ backgrounds. In this section we will construct the most general $\la$YB deformations of such spaces by presenting a detailed construction for the spheres $S^N=SO(N+1)/SO(N)$ and extending the results to the AdS factors by making the appropriate analytic continuations.

\bigskip

As demonstrated in section \ref{SecRealR}, the most general real solution of the mCYBE equation for a real group is given by the canonical construction (\ref{mCYBcan}), so deformations of $SO(N+1)/SO(N)$ are fully characterized by the orientation of the $SO(N)$ subgroup relative to the root system used in (\ref{mCYBcan}). The vast majority of such orientations lead to $R$--matrices which don't satisfy the constraints (\ref{YBcosetConstr}), so the resulting deformations are not integrable. Nevertheless, some choices of $SO(N)$ do lead to nontrivial integrable deformations, and several examples have been discussed in the past \cite{GenLambda,ChLGL}. In all these cases the generalized lambda--deformation was reducible to the standard version, so one may wonder whether this is a general feature. In this section we will demonstrate that the reduction to the standard deformation is a peculiarity associated with small groups, and we will construct the most general deformation of a coset for arbitrary $N$. Furthermore, in the most interesting case of $S^5$, we will find a six--parameter family of integrable models which are not equivalent to the standard lambda--deformation. 

\bigskip

\noindent
{\bf Construction of the R--matrix}

To construct the most general integrable deformation of $SO(N+1)/SO(N)$, we begin with choosing a specific root system in $SO(N+1)$ and defining the corresponding canonical R--matrix (\ref{mCYBcan}),
which will be denoted by $\RR$ (recall the map (\ref{RmatrReadOff})). As demonstrated in section \ref{SecRealR}, any real antisymmetric solution of the mCYBE is equivalent to (\ref{mCYBcan}), so it can be constructed by rotating $\RR$:
\bea\label{RRrotG}
{\RR}_g=g {\RR}g^{-1}, \quad g\in SO(N+1)
\eea
Next we select a specific subgroup $SO(N)\subset SO(N+1)$ and identify the elements $g$ which lead to the matrix ${\RR}_g$ satisfying the constraint (\ref{YBcosetConstr}). Once all such R--matrices are found, the construction reviewed in section \ref{SectReview} leads to an integrable sigma--model for every allowed 
${\RR}_g$. 

Let us define (anti--hermitian) generators of $G=SO(N+1)$ in the fundamental representation:
\bea\label{genFundMain}
(T_{mn})_{ab}=\delta_{ma}\delta_{nb}-\delta_{mb}\delta_{na},\quad (m,n)={1,\dots,(N+1)},
\eea
and choose the subgroup $F=SO(N)$ to be generated by (\ref{genFundMain}) with $(m>1,n>1)$.
The operators (\ref{genFundMain}) satisfy the standard commutation relations for $G$, which hold in all representations:
\bea\label{SOalgebraMain}
[T_{mn},T_{pq}]=\delta_{np}T_{mq}-\delta_{mp}T_{nq}-\delta_{nq}T_{mp}+\delta_{mq}T_{np}. 
\eea
The canonical R--matrix is constructed in the Appendix \ref{AppRmatr}, and its nonzero elements are given by 
(\ref{RmatrCanon}):
\bea\label{RmatrCanonMain}
\begin{array}{c}\RR T_{p,q}=-T_{p+1,q}\\
\RR T_{p,q+1}=-T_{p+1,q+1}
\end{array},\quad
\begin{array}{c}
\RR T_{p+1,q}=T_{p,q}\\
{\RR}T_{p+1,q+1}=T_{p,q+1}
\end{array},\quad {\RR}T_{p,p+1}=\sum_{odd\ s} R_{sq}T_{s,s+1}.
\eea
Here  $(p,q)$ are odd integers and $p<q$. For even values of $N$, matrix $\RR$ has an additional set of nontrivial elements:
\bea\label{RmatrCanonMain2}
\RR T_{p,N+1}=T_{p+1,N+1},\quad  \RR T_{p+1,N+1}=-T_{p,N+1}.
\eea
The canonical solution (\ref{RmatrCanonMain})--(\ref{RmatrCanonMain2}) of the Yang--Baxter equation satisfies the constraint (\ref{MatrixConstr}), but this condition is violated by a generic rotation 
(\ref{RRrotG}). The detailed analysis presented in the Appendix \ref{AppRmatr} shows that the constraint (\ref{MatrixConstr}) is preserved by $\RR_g$ if and only if $g$ is an element of the subgroup $F$:
\bea\label{GeneralCanonic}
{\RR}\rightarrow {\RR}_f=f \RR f^{-1},\qquad f\in F.
\eea
Let us now construct the resulting integrable deformation of the sigma--model. We will mostly focus on the odd values of $N$ and discuss the minor modification associated with even values on page \pageref{EvenNpage}.

The deformation for odd values of $N$ will be constructed in two steps: we will begin with evaluating the metric corresponding to the canonical 
R--matrix $\RR$, and then extend the result to the general case (\ref{GeneralCanonic}). 
\bigskip

\noindent
{\bf Deformation for the canonical R--matrix.}

The deformed sigma model is described by the action (\ref{GenLambdaAction}), which leads to the metric
\bea\label{DeformMetr}
 ds^2=\frac{k}{4\pi}L^T(\hat{\lambda}^{-1}-D)^{-1}
 \Big[\hat{\lambda}^{-1}\hat{\lambda}^{-T}-I\Big](\hat{\lambda}^{-1}-D)^{-T}L.
 \eea 
Using Greek letters $(\alpha,\dots)$ to denote the coset and Latin indices $(a,\dots)$ to label the subgroup, we conclude that the matrix $\hat{\lambda}^{-1}$ defined by (\ref{lambdaMtrDef}) has a block structure
 \bea\label{DefLaInv}
 {\hat\lambda}^{-1}=\begin{pmatrix}
 	H_{\alpha\beta}&0\\
 	0&\delta_{ab}
 \end{pmatrix}, \quad H_{\alpha\beta}=(I+E)_{\alpha\beta}.
 \eea 
Substitution of these expressions into (\ref{DeformMetr}) leads to a compact expression for the metric\footnote{Some details of the transition from equation (\ref{DeformMetr}) to the final result (\ref{ds2ProdStr})--(\ref{QalBetAsJH}) can be found in the Appendix \ref{AppJJ}.}
\bea\label{ds2ProdStr}
  ds^2=\frac{k}{4\pi}e_0^{\alpha}Q_{\alpha\beta} e_0^{\beta},
\eea
Here
\bea\label{ds2ProdStr2}
(e_0)_\alpha=\Big[(D_{ab}-\delta_{ab}\big)^{-1}D_{a\alpha}\Big]^T L_b-L_\alpha
\eea
are the original frames, and matrix $Q_{\alpha\beta}$ contains all information about the deformation. The direct evaluation outlined in the Appendix \ref{AppJJ} gives
\bea\label{QalBetAsJH}
Q_{\alpha\beta}&=&\left\{I-\Big[(1-J^{-1}H)^{-1}+(1-J^{-1}H)^{-T}\Big]\right\}_{\alpha\beta},\\
J_{\alpha\beta}&=&D_{\alpha\beta}-D_{\alpha b}(D_{ab}-\delta_{ab})^{-1}D_{a\beta}.\nonumber
\eea
Note that the frames $(e_0)_\alpha$ and the matrix $Q_{\alpha\beta}$ are invariant under diffeomorphisms, 
so there is a one--to--one correspondence between matrices $Q$ and distinct deformations. 

As demonstrated in \cite{ChLGL}, every coset admits a ``canonical'' gauge, where matrix $Q$ has only constant entries. In particular, for $SO(N+1)/SO(N)$ such canonical parameterization is given by (\ref{gElemApp}), and the matrix $J_{\alpha\beta}$ becomes
\bea\label{Jidentity}
J_{\alpha\beta}=(-1)^{1+\alpha}\delta_{\alpha\beta}.
\eea
This relation was introduced and verified in \cite{DSTLmbAdS5} for $N=3,4,5$, and in the Appendix \ref{AppJJ} we present a recursive proof for all values of $N$.

To complete the construction of $Q_{\alpha\beta}$, we need to simplify the expression (\ref{DefLaInv}) for 
$H_{\alpha\beta}$. Using the relation (\ref{EmatrYB}), we find
\bea\label{HalBet}
 H_{\alpha\beta}=\Big[I+\frac{1}{t}(1-\eta\RR)^{-1}\Big]_{\alpha\beta}\,.
\eea 
For the canonical R--matrix (\ref{RmatrCanonMain}), it is convenient to choose a particular order of the generators\footnote{Recall that here we are focusing on odd values of $N=2k-1$, and the minor modifications for even $N$ will be discussed below.}
\bea\label{Generators1}
&&T_\alpha=\{T_{12};T_{13},\dots T_{1,(N+1)}\},\\ 
&&T_a=\{T_{34},T_{56},\dots,T_{N,(N+1)};T_{23},T_{24},\dots T_{2,(N+1)};
T_{35},T_{45},T_{36},T_{46};\dots\}.\nonumber
\eea
The last set continues with blocks $\{T_{p,q},T_{p+1,q},T_{p,q+1},T_{p+1,q+1}\}$ for all odd values 
of $p$ and $q$. In the basis $(T_\alpha,T_a)$, the canonical R--matrix  (\ref{RmatrCanonMain}) has the form
\bea\label{RmatrInBasis}
\RR=\left[
\begin{array}{cc|ccc}
0&0&V&0&0\\
0&0&0&-I_{N-1}&0\\
\hline
-V^T&0&A_{k-1}&0&0\\
0&I_{N-1}&0&0&0\\
0&0&0&0&[\otimes\,\eps]^{(k-1)(k-2)}
\end{array}
\right]\,,
\eea 
where $\eps$ is an antisymmetric $2\times 2$ matrix with two nontrivial entries: $\eps_{12}=-\eps_{21}=-1$. An antisymmetric matrix $A_{k-1}$ and a vector $V$ form the elements $R_{pq}$ of the R--matrix acting on the Cartan subalgebra (see (\ref{RmatrCanonMain})). To evaluate $H_{\alpha\beta}$ we need the upper left corner of the matrix $(1-\eta\RR)^{-1}$, and it is clear that the answer is 
\bea\label{RinvInterm}
\left[(1-\eta\RR)^{-1}\right]_{\alpha\beta}=
\left[
\begin{array}{cc}
\frac{1}{1+\gamma\eta^2}&0\\
0&\frac{1}{1+\eta^2}I_{N-1}
\end{array}
\right],\qquad \gamma=V^T(1-\eta A)^{-1} V\,.
\eea
We used the standard block inversion formula:
\bea
\left\{\left[I-\begin{pmatrix}
 	0&V_1\\
 	V_2^T&A
 \end{pmatrix}\right]^{-1}\right\}_{11}=\left[1-V^T_2(1-A)^{-1}V_1\right]^{-1}\,.\nonumber
\eea
Expression (\ref{RinvInterm}) leads to the final answer for $H_{\alpha\beta}$ corresponding to the canonical R--matrix:
\bea\label{HalBetCan}
H_{\alpha\beta}=
\left[
\begin{array}{cc}
1+\frac{1}{t}\frac{1}{1+\gamma\eta^2}&0\\
0&\left[1+\frac{1}{t}\frac{1}{1+\eta^2}\right]I_{N-1}
\end{array}
\right]\equiv \left[
\begin{array}{cc}
\frac{\theta}{\la}&0\\
0&\frac{1}{\la}I_{N-1}
\end{array}
\right]\,.
\eea
The standard $\la$--deformation \cite{SftsGr,MoreLmb,DSTLmbAdS5} is recovered by setting $\eta=0$, then $H$ is proportional to the unit matrix:
\be
H_{\alpha\beta}=\lambda^{-1}\delta_{\alpha\beta}\,,\quad \la^{-1}=1+\frac{1}{t}\,.\nonumber
\ee
The information about solution of the Yang--Baxter equation is encoded in a single parameter $\theta$ entering (\ref{HalBetCan}), and the relevant $H_{11}$ corresponds to the generator in the overlap of the coset and the Cartan subalgebra.  Since both $H_{\alpha\beta}$ and $J_{\alpha\beta}$ are diagonal, the expression for the deformed metric is rather simple:
\bea\label{AnsCanNodd}
ds^2=\frac{k}{4\pi}Q_{\alpha\beta} e_0^\alpha e_0^\beta,\quad 
Q=\text{diag}\{\frac{1}{\nu},\mu,\frac{1}{\mu},\dots,\mu,\frac{1}{\mu}\},\quad
\mu=\frac{1-\lambda}{1+\lambda},\quad
\nu=\frac{\theta-\lambda}{\theta+\lambda}\,.
\eea
Thus the canonical R--matrix leads to the deformation (\ref{AnsCanNodd}) parameterized by two constants, $\la$ and $\theta$. Let us now extend this result to the general R--matrix (\ref{GeneralCanonic}). 
\bigskip

\noindent
{\bf Deformation for the general R--matrix.}

An extension of the expression (\ref{AnsCanNodd}) to the R--matrix (\ref{GeneralCanonic}) follows the logic outlined above, but it involves some long algebraic manipulations. The readers interested only in the final answer can go directly to equation (\ref{QOddFinal}).

\bigskip

The matrix $Q$ is still computed using the relation (\ref{QalBetAsJH}), but $H_{\alpha\beta}$ is longer given by the equation (\ref{HalBetCan}). This complicates the procedure for constructing matrix $Q$ and leads to new deformation parameters. As we have already demonstrated, the most general R--matrix leading to integrable $\la$YB deformation is given by (\ref{GeneralCanonic}). Action by an element $f\in SO(N)$ does not mix the generators of the subgroup and the coset, so in the basis (\ref{RmatrInBasis}), $f$ has the form
\bea
f=\left[
\begin{array}{c|c}
f_1&0\\
\hline
0&f_2
\end{array}
\right]\,.
\eea
Then the transformation (\ref{GeneralCanonic}) leads to rotation of the matrix $H_{\alpha\beta}$  defined by (\ref{HalBet}):
\bea\label{Eqn420}
[{\hat\lambda}^{-1}]_{\alpha\beta}=H^{(f)}_{\alpha\beta}=[f_1 H f_1^{-1}]_{\alpha\beta},
\eea
and substitution into (\ref{QalBetAsJH}) gives
\bea\label{PalBetCanRotXX}
Q_{\alpha\beta}&=&\left\{I-P^{-1}J-JP^{-T}\right\}_{\alpha\beta},\quad 
P_{\alpha\beta}\equiv [J-f_1 H f_1^{-1}]_{\alpha\beta}\,.
\eea
Recalling the expression (\ref{HalBetCan}) for the matrix $H_{\alpha\beta}$, we find
\bea\label{PalBetCanRot}
P=J-\frac{1}{\la}I-f_1^{-1}
\left[
\begin{array}{cc}
\frac{\theta-1}{\la}&0\\
0&0_{N-1}
\end{array}
\right]f_1\,.
\eea
According to the order (\ref{Generators1}), the non--zero eigenvalue in the last term corresponds to the generator $T_{12}$. For $f_1=I$ equations (\ref{PalBetCanRotXX})--(\ref{PalBetCanRot}) recover the earlier result (\ref{AnsCanNodd}). In general, matrix $f_1$ appearing in (\ref{PalBetCanRot}) is an element of the fundamental representation of $SO(N)$, and it is convenient to use parameterization introduced in \cite{Bars,DSTLmbAdS5}:
\bea\label{f1Param}
f_1=\left[
\begin{array}{cc}
1&0\\
0&h
\end{array}
\right]\left[
\begin{array}{cc}
b-1&b X^T\\
-b X&1-b X X^T
\end{array}
\right],\qquad b=\frac{2}{1+X^T X}\,.
\eea
Here $h$ is an element of $SO(N)/SO(N-1)$ coset, and $X$ is an $(N-1)$--component column vector.
Then the parameters associated with the coset $h$ cancel in (\ref{PalBetCanRot}), leading to the expression for the matrix $P$:
\bea\label{PmatrFin1}
P=J-\frac{1}{\la}I-\frac{\theta-1}{\la}
\left[
\begin{array}{cc}
(b-1)^2&b(1-b)X^T\\
b(1-b)X&b^2 XX^T
\end{array}
\right]\,.
\eea
Although there is a one--to--one correspondence between inequivalent matrices $P$ and parameters 
$(\la,\theta,X)$ appearing in the last expression, to evaluate matrix $Q$ it is convenient to use another parameterization that has some redundancy. Specifically, we introduce rotations
$S\in SO(\frac{N-1}{2})\times SO(\frac{N-1}{2})$ that act on two spaces\footnote{To agree with matrix indices of $P$, we assume that $i$ in $X_i$ takes values $\{2,3,\dots N\}$.}: $(X_2,X_4,\dots X_{N-1})$ and $(X_3,X_5,\dots,X_N)$. Using such transformation, the matrix $P$ can be written as
\bea
P=S^{-1}{\tilde P}S,
\eea
where ${\tilde P}$ has the form (\ref{PmatrFin1}) with only two nontrivial components of $X$: $(X_2,X_3)$. Since rotation $S$ commutes with $J$, the matrix $Q$ can be rewritten as
\bea
Q=S^{-1}\left[I-{\tilde P}^{-1}J-J{\tilde P}^{-1}\right]S\equiv S^{-1}{\tilde Q}S
\eea
Recalling the result (\ref{AnsCanNodd}) for $\theta=1$, we conclude that 
\bea
{\tilde Q}={\hat Q}+\text{diag}\{0,0,0,\mu,\frac{1}{\mu},\dots,\mu,\frac{1}{\mu}\},
\eea
with nonzero ${\hat Q}_{\alpha\beta}$ only for $(\alpha,\beta)=\{1,2,3\}$. Explicit evaluation of the relevant block gives
\bea
{\hat Q}=q\left[\begin{array}{ccc}
(z-1)(1+\la)+2\la z(b^2 X_3^2-1)&0&2(1-b)b\la X_3 z\\
0&\frac{(\la-1)^2(z-1)}{\la+1}&0\\
2(1-b)b\la X_3 z&0&(\la+1)(z-1)-2\la bz(2-b)
\end{array}\right]\nonumber
\eea
Here we defined two parameters, $ z=\frac{\theta-1}{\la-1}$ and $q$, and the expression for the latter is given below. 
Additional rotation in the $(1,3)$ plane brings the $Q$--matrix to the final form:
\bea\label{QOddFinal}
Q&=&S^{-1}\text{diag}\{-(\la+\theta)q,\frac{(1-\la)(\la-\theta)}{\la+1}q,
\frac{1}{\mu},\mu,\frac{1}{\mu},\dots,\mu,\frac{1}{\mu}\} S\nn
q&=&\frac{\la+1}{(\la+1)(\la-\theta)+2\la \sigma (\theta-1)},\quad
\sigma=b^2 X_2^2,\quad \mu=\frac{1-\lambda}{1+\lambda}\,.
\eea
Here $S$ is an arbitrary $SO(\frac{N+1}{2})\times SO(\frac{N-1}{2})$ rotation that commutes with 
matrix $J$. Note that parameters $(X_2,X_3)$ enter only in one combination
\bea
\sigma=b^2 X_2^2=\frac{4 X_2^2}{(1+X_2^2+X_3^2)^2},\quad 0\le \sigma\le 1.
\eea
Different rotations $S$ may lead to the same matrix $Q$, and for generic values of 
$(\la,\theta,\sigma)$ this redundancy can be removed by focusing on\footnote{Parameterization used here is inspired by (\ref{f1Param}).}
\bea\label{RforQodd}
&&\hskip -1cm
S\in\frac{SO(\frac{N+1}{2})}{SO(\frac{N-1}{2})}\times \frac{SO(\frac{N-1}{2})}{SO(\frac{N-3}{2})}:\quad
S=I+\alpha Y_+^T Y_-+\beta Z_+^T Z_--2(Y_0^T Y_0+Z_0^T Z_0),\nn
&&\hskip -1cm\begin{array}{ll}
Y_\pm=(\pm 1,0,Y_1,0,\dots, 0,Y_k)\\
Z_\pm=(0,\pm 1,0,Z_1,\dots, Z_{k-1},0)
\end{array}\
\begin{array}{ll}
Y_0=(1,0,0,\dots, 0)\\
Z_0=(0,1,0,\dots, 0)
\end{array},\
\alpha=\frac{2}{Y_+ Y_+^T},\ \beta=\frac{2}{Z_+ Z_+^T}
\quad 
\eea
Thus matrix $Q$ is specified by $N+1$ parameters: $(\la,\theta,\sigma)$ and two vectors, $(Y_1\dots Y_k)$ and $(Z_1\dots Z_{k-1})$. This should be contrasted with the regular $\la$--deformation, which is completely determined by $\la$: to recover this result one should set $\theta=1$ in (\ref{QOddFinal}) and observe that matrix $S$ cancels in $Q$. Equations (\ref{QOddFinal}) and (\ref{RforQodd}) along with (\ref{ds2ProdStr}) constitute the final answer for the most general $\la$YB deformation. 

\bigskip

Let us discuss two special cases of (\ref{QOddFinal}). For $\sigma=0$ one finds a rotation of the 
canonical result (\ref{AnsCanNodd}):
\bea
Q_{\sigma=0}&=&S^{-1}\text{diag}\{\frac{\theta+\la}{\theta-\la},\mu,\frac{1}{\mu},\mu,\dots,\frac{1}{\mu},\mu\} S,\nonumber
\eea
and transformation associated with vector $Z$ cancel in this expression. Thus there are only $\frac{N-1}{2}$ 
parameters in addition to $(\la,\theta)$. 
Another interesting limit, $\sigma=1$, gives
\bea\label{Qsigma1}
Q_{\sigma=1}&=&S^{-1}\text{diag}\{\frac{1}{\mu},\frac{\la-\theta}{\la+\theta},\frac{1}{\mu},\mu,\dots,\frac{1}{\mu},\mu\} S,
\eea
and now it is the $Y$--parameters that become irrelevant. In particular, for $N=3$, all parameters associated with rotation disappear, and furthermore, the parameter $\theta$ can be removed as well since the matrix 
$Q_{\sigma=1}$ can be rewritten as
\bea
Q_{\sigma=1}=\text{diag}\{\frac{1}{\mu},\nu,\frac{1}{\mu}\}=\sqrt{\frac{\nu}{\mu}}\text{diag}\{
\frac{1}{\tilde\mu},\tilde\mu,\frac{1}{\tilde\mu }\}\,.
\eea
Absorbing the prefactor into the coupling constant $k$ of the sigma--model, one concludes that the generalized lambda--deformation of $S^3$ with $\sigma=1$ is equivalent to the standard one \cite{ChLGL}. As one can see from (\ref{Qsigma1}), this peculiarity disappears for $N>3$.

\bigskip
\noindent
{\bf Deformation for even values of $N$}
\label{EvenNpage}

Let us briefly discuss the minor modifications of the previous construction that appear for even values of $N=2(k+1)$. 
For the deformation corresponding to the canonical R--matrix, the expression (\ref{AnsCanNodd}) is replaced by 
\bea\label{AnsCanNeven}
ds^2=\frac{k}{4\pi}Q_{\alpha\beta} e_0^\alpha e_0^\beta,\quad 
Q=\text{diag}\{\frac{1}{\nu},\mu,\frac{1}{\mu},\dots,\mu\},\quad
\mu=\frac{1-\lambda}{1+\lambda},\quad
\nu=\frac{\theta-\lambda}{\theta+\lambda}\,.
\eea
In particular, for the standard lambda deformation ($\theta=1$) the eigenvalues $\mu$ and 
$\frac{1}{\mu}$ have the same degeneracies. 

The most general deformation (\ref{QOddFinal}) is replaced by 
\bea\label{QEvenFinal}
Q&=&S^{-1}\text{diag}\{-(\la+\theta)q,\frac{(1-\la)(\la-\theta)}{\la+1}q,
\frac{1}{\mu},\mu,\frac{1}{\mu},\dots,\mu\} S\,,\nn
q&=&\frac{\la+1}{(\la+1)(\la-\theta)+2\la \sigma (\theta-1)},\quad
\mu=\frac{1-\lambda}{1+\lambda}\,,
\eea
where $S$ is a product of two identical cosets:
\bea\label{RforQeven}
&&\hskip -1cm
S\in\frac{SO(\frac{N}{2})}{SO(\frac{N-2}{2})}\times \frac{SO(\frac{N}{2})}{SO(\frac{N-2}{2})}:\quad
S=I+\alpha Y_+^T Y_-+\beta Z_+^T Z_--2(Y_0^T Y_0+Z_0^T Z_0),\nn
&&\hskip -1cm\begin{array}{ll}
Y_\pm=(\pm 1,0,Y_1,0,\dots, Y_k,0)\\
Z_\pm=(0,\pm 1,0,Z_1,\dots, 0,Z_{k})
\end{array}\
\begin{array}{ll}
Y_0=(1,0,0,\dots, 0)\\
Z_0=(0,1,0,\dots, 0)
\end{array}\
\alpha=\frac{2}{Y_+ Y_+^T},\ \beta=\frac{2}{Z_+ Z_+^T}
\quad 
\eea
Again, the matrix $Q$ is specified by $(N+1)$ parameters: $(\la,\theta,\sigma)$ and two vectors with the same length: $(Y_1\dots Y_k)$ and $(Z_1\dots Z_{k})$. Although algebraically the deformations for odd and even $N$ are very similar, the construction with odd $N$ is much more interesting for embedding the integrable backgrounds in string theory. 

\bigskip

Let us summarize the results of this rather technical section. We have constructed the most general $\la$YB deformations of the spheres $S^N$, and the results are given by (\ref{QOddFinal})--(\ref{RforQodd}) and (\ref{QEvenFinal})--(\ref{RforQeven}). The explicit expressions for the undeformed frames $e_0^\alpha$ can be found in \cite{DSTLmbAdS5}, and they are rather complicated. The $\la$YB--deformed AdS space can be obtained by a simple analytic continuation of frames, as it was done for the standard lambda--deformation in \cite{DSTLmbAdS5}. As demonstrated in the Appendix \ref{AppBzero}, the deformed solutions have vanishing Kalb--Ramond field, and according to (\ref{dilaton}), (\ref{HJdilat}), the dilaton is not modified by the deformation.

\section{Summary} 

In this article we have explored the analytical structure of the generalized lambda deformations ($\la$YB) introduced in \cite{GenLambda} and found the explicit solutions leading to integrable deformations of spheres and AdS spaces. Specifically, we have demonstrated that the reality of the R--matrix combined with the integrability constraint coming from the coset projection (\ref{YBcosetConstr}) imply that the 
$\la$YB deformation of $S^N$ introduces $(N+1)$ free parameters.   
Furthermore, we constructed the resulting metrics, which are given by (\ref{ds2ProdStr}), (\ref{QOddFinal})--(\ref{RforQeven}), (\ref{QEvenFinal})--(\ref{RforQeven}). It would be interesting to explore the physical properties of the new geometries by studying propagation of probe particles and fundamental strings.


\section*{Acknowledgments}

We thank Yuri Chervonyi for discussions. This work was supported by the DOE grant 
DE\,-\,SC0017962.

\appendix

\section{R--matrix for the $SO(N+1)/SO(N)$ coset}
\renewcommand{\theequation}{A.\arabic{equation}}
\setcounter{equation}{0}
\label{AppRmatr}

In this appendix we will outline the derivation of the canonical R--matrix (\ref{RmatrCanonMain}) and demonstrate that the rotation (\ref{RRrotG}) preserves the projection (\ref{YBcosetConstr}) if and only if $g$ is an element of the subgroup $F$. Since the most interesting integrable string theories corresponds to the backgrounds of the form AdS$_p\times$S$^p$, here we will focus on cosets describing spheres $S^N=SO(N+1)/SO(N)$, although the final result is expected to hold for general $F/G$. The deformations of the AdS factors can be obtained by an analytic continuation of the deformed $S^N$.

\bigskip

We begin with choosing (anti--hermitian) generators of $G=SO(N+1)$ in the fundamental representation:
\bea\label{genFund}
(T_{mn})_{ab}=\delta_{ma}\delta_{nb}-\delta_{mb}\delta_{na},\quad (m,n)={1,\dots,(N+1)},
\eea
These generators satisfy the standard commutation relations for $G$, which hold in all representations:
\bea\label{SOalgebra}
[T_{mn},T_{pq}]=\delta_{np}T_{mq}-\delta_{mp}T_{nq}-\delta_{nq}T_{mp}+\delta_{mq}T_{np}. 
\eea
We further choose the subgroup $F=SO(N)$ to be generated by (\ref{genFund}) with $(m,n)={2,\dots,(N+1)}$. 

The goal of this appendix is to find R--matrices satisfying the Yang--Baxter equation (\ref{myb}) and the constraint  (\ref{constraint}). The R--matrix acting on the generators $T_{mn}$ has two composite indices $\RR_{(mn),(pq)}$:
\bea
\RR T_{pq}\equiv \sum_{m,n}\RR_{(mn),(pq)}T_{mn},
\eea
and the constraint (\ref{constraint}) becomes
\bea\label{MatrixConstr}
\RR_{(1,m),(1,n)}=0.
\eea
All anti--symmetric solutions of the modified Classical Yang--Baxter equation (\ref{myb}), mCYBE,  have been classified \cite{BelavDrinf}, and the results are summarized on page \pageref{BelDrinClassPage}. In this article we are interested in real R--matrices for real algebras, then the classification collapses to the canonical solution (\ref{mCYBcan}). The resulting R--matrix acts on the Cartan generators $H_i$ and the roots $E_\alpha$ as 
\bea 
\label{CanSoln}
\RR H_i=R_{ij}H_j,\quad \RR E_{\alpha}=-i E_\alpha,\quad \RR E_{-\alpha}=i E_{-\alpha}\,,
\eea
where $R_{ij}$ is an arbitrary anti--symmetric matrix. For future reference we recall the commutation relations involving the roots and the Cartan subalgebra:
\bea
[H_i,H_j]=0,\quad [H_i,E_\alpha]=\alpha_i E_\alpha,\quad 
[E_\alpha,E_\beta]=e_{\alpha,\beta}E_{\alpha+\beta},\quad
[E_\alpha,E_{-\alpha}]=\sum_i {\tilde\alpha}^i H_i\,.
\eea
The explicit expressions for the roots are different for odd and even $N$, so we will begin with a detailed analysis for $N=2k-1$, and discuss the extension to the even values of $N$ after equation (\ref{RmatrCanon}). 

For the odd values of $N=2k-1$, we choose the Cartan subalgebra generated by
\be 
\{T_{12},T_{34},\dots,T_{N,N+1}\}.
\ee 
Then the roots have the structure 
\be 
\label{root}
\alpha=\{(a_1,a_2,\dots,a_k)\}\,,
\ee 
where $(k-2)$ components of $a_i$ vanish and the remaining two components are equal to $\pm i$.  
The `raising' ladder operators corresponding to the positive roots are characterized by two odd numbers $(p,q)$ with $p<q$ and by an additional index taking two values $\pm$:
\bea
E^{(+)}_{pq}&=&T_{p,q}+iT_{p+1,q}+iT_{p,q+1}-T_{p+1,q+1}\nn
E^{(-)}_{pq}&=&T_{p,q}+iT_{p+1,q}-iT_{p,q+1}+T_{p+1,q+1}
\eea
The non--zero commutators involving the roots and the Cartan generators are
\bea
[T_{p,p+1},E^{(\pm)}_{p,q}]=iE^{(\pm)}_{p,q},\quad 
[T_{q,q+1},E^{(\pm)}_{p,q}]=\pm iE^{(\pm)}_{p,q}
\eea
and the positivity of the root corresponding to $E^{(\pm)}_{p,q}$ is reflected in the sign in the right--hand side of the first relation. 

By definition (\ref{CanSoln}), the canonical R--matrix acts as
\bea
{\r}E^{(\pm)}_{p,q}=-iE^{(\pm)}_{p,q},\quad
{\r}{\overline{E^{(\pm)}_{p,q}}}=i{\overline{E^{(\pm)}_{p,q}}},\quad
{\r}T_{q,q+1}=\sum_{odd\ s} R_{sq}T_{s,s+1}.
\eea
In other words, for odd $(p,q)$ ($p<q$) we find
\bea\label{RmatrCanon}
\begin{array}{c}{\r}T_{p,q}=-T_{p+1,q}\\
{\r}T_{p,q+1}=-T_{p+1,q+1}
\end{array},\quad
\begin{array}{c}
{\r}T_{p+1,q}=T_{p,q}\\
{\r}T_{p+1,q+1}=T_{p,q+1}
\end{array},\quad {\r}T_{p,p+1}=\sum_{odd\ s} R_{sq}T_{s,s+1}.
\eea
To extend this result to even values of $N$, we observe that in this case the Cartan subalgebra is generated by
\be 
\{T_{12},T_{34},\dots,T_{N-1,N}\}\,,
\ee 
and the roots still have the structure (\ref{root}), although now $a_i$ can have either one or two non-zero components with values $\pm i$. The canonical solution (\ref{RmatrCanon}) should be supplemented by 
\bea\label{RmatrCanonX}
\RR T_{p,N+1}=T_{p+1,N+1},\quad  \RR T_{p+1,N+1}=-T_{p,N+1}.
\eea
The arguments presented in the rest of this appendix are equally applicable to odd and even values of $N$.

The canonical solution (\ref{RmatrCanon}) of the Yang--Baxter equation satisfies the constraint (\ref{MatrixConstr}). Moreover, rotation of the R--matrix by an arbitrary element $g\in SO(N+1)$,
\bea
\r\rightarrow \RR_g=g\RR g^{-1}\,,
\eea
does not affect the mCYB equation, so we only need to ensure that the constraint (\ref{MatrixConstr}) is preserved under such rotation. The rest of this appendix is dedicated to extraction of restrictions on $g$. 

\bigskip

We begin by observing that any element $g$ of the group $G$ can be written as a product 
\bea\label{Nov6a}
g=f q,\quad f\in F,\quad q\in G/F,
\eea
and for $SO(N+1)/SO(N)$ coset we further specify the gauge:
\bea\label{Nov6b}
q=h\, f_2 f_3,\qquad
h=e^{i\theta_1 T_{1,2}},\quad f_2=e^{i\theta_2 T_{2,3}},\quad 
f_3=e^{i\theta_3 T_{3,4}}}\dots {e^{i\theta_N T_{N,N+1}}.
\eea
The constraint (\ref{MatrixConstr}) applied to ${\r}_g$ gives
\bea\label{RgConstr}
0=\mbox{Tr}[T_{1,m} {\r}_g T_{1,n}]=\mbox{Tr}[(f^{-1}T_{1,m}f) {\r}_q (f^{-1}T_{1,n}f)]\,.
\eea
Since the rotated generator $f^{-1}T_{1,m}f$ is a linear combination of $T_{1,s}$\footnote{We also note that transformation between sets $\{f^{-1}T_{1,m}f\}$ and $\{T_{1,s}\}$ is invertible.}, the last constraint is equivalent to 
\bea
\mbox{Tr}[T_{1,m}{\r}_q T_{1,n}]=0\,.
\eea
This relation has to hold for all $(m,n)$, in particular, for $(m,n)=(3,2)$ we find
\bea\label{Constr1}
0=\mbox{Tr}[T_{1,3}{\r}_q T_{1,2}]=
\mbox{Tr}[q^{-1}T_{1,3}q{\r} q^{-1}T_{1,2}q]\,.
\eea
Observing that $f_3$ commutes with $T_{12}$ and that $f_3^{-1}T_{13}f_3$ is a linear combination of $T_{1,m}$, we find
\bea
{\r}[q^{-1}T_{1,2}q]&=&{\r}[f_3^{-1}(\cos\theta_2 T_{1,2}+i\sin\theta_2 T_{1,3})f_3]
\nn
&=&\cos\theta_2\sum_{odd~p}R_{1,p}T_{p,p+1}+i\sin\theta_2 f_3^{-1} T_{2,3}f_3.
\eea
Here we used (\ref{RmatrCanon}) and (\ref{RmatrCanonX}). Substitution of the last equation into (\ref{Constr1}) gives
\bea
\cos\theta_2 \sum_{odd~p>1}R_{1,p}\mbox{Tr}[q^{-1}T_{1,3}q T_{p,p+1}]+
i\sin\theta_2 \mbox{Tr}[h^{-1}T_{1,3}h T_{2,3}]=0.
\eea
Observing that $q^{-1}T_{1,3}q$ is a linear combination of $T_{1,m}$, we conclude that the first term in the last expression vanishes, and the constraint (\ref{Constr1}) takes the final form
\bea
\mbox{Tr}[T_{1,3}{\r}_q T_{1,2}]=-\sin\theta_2\sin\theta_1=0.
\eea
This leads to two options:
\bea
\theta_1=0:&&q=f_2f_3\in F\nn
\theta_2=0:&&q=hf_3=f_3 h.
\eea
Recalling the notation (\ref{Nov6a}), (\ref{Nov6b}), we conclude that both options can be summarized as
\bea\label{Nov6c}
g=f e^{i\theta_1 T_{1,2}}\equiv fh,\quad f\in F.
\eea
Thus we have demonstrated that if ${\cal R}_g$ satisfies the constraint (\ref{RgConstr}), then $g$ must have the form (\ref{Nov6c}). The explicit expressions (\ref{RmatrCanon}) and (\ref{RmatrCanonX}) imply that $h$ commutes with ${\RR}$, then
\bea
{\RR}_g=fh {\RR}h^{-1}f^{-1}={\RR}_f.
\eea
The last equation constitutes the main result of this appendix.

\bigskip

To summarize, we have demonstrated that the rotation (\ref{RRrotG}) of the canonical R--matrix preserves the projection (\ref{YBcosetConstr}) if and only if $g$ is an element of the subgroup $F$. This result is used in section \ref{SectMainDeform} to construct the most general $\la$YB deformation of the coset $S^N$.



 \section{Properties of the matrix $J_{\alpha\beta}$}
 \renewcommand{\theequation}{B.\arabic{equation}}
\setcounter{equation}{0}
\label{AppJJ}

In this appendix we discuss some properties of matrix $J$ used in section \ref{SectMainDeform}. Specifically, we derive equations (\ref{QalBetAsJH}) and prove the identity (\ref{Jidentity}). Some special cases of the formulas derived in this appendix were presented in \cite{DSTLmbAdS5}.

\bigskip

We begin with outlining the details of the transition from equation (\ref{DeformMetr}) to the final expression for the metric (\ref{ds2ProdStr})--(\ref{QalBetAsJH}). Expression (\ref{DeformMetr}) contains the inverse of the matrix 
\bea\label{MatrM}
M=(\hat{\lambda}^{-1}-D)^{T}\equiv\left[\begin{array}{cc}
{\bf D}&{\bf C}\\
{\bf B}&{\bf A}
\end{array}\right],
\eea
where the block ${\bf A}$ corresponds to the subgroup and the block ${\bf D}$ corresponds to the coset. 
Recall that matrices $D$ and ${\hat\lambda}^{-1}$ are defined by
\bea
D_{MN}=\text{Tr}[T_M g T_N g^{-1}],\qquad {\hat\lambda}^{-1}=\left[\begin{array}{cc}
 	H_{\alpha\beta}&0\\
 	0&\delta_{ab}
 \end{array}\right]\,.
\eea
The inverse of the matrix (\ref{MatrM}) can be evaluated using the method introduced in \cite{DSTLmbAdS5}:
\bea
M^{-1}=
\left[\begin{array}{cc}
{\bf P}^{-1}&0\\
0&{\bf I}
\end{array}\right]\left[\begin{array}{cc}
{\bf I}&-{\bf C}{\bf A}^{-1}\\
-{\bf A}^{-1}{\bf B}{\bf P}^{-1}&{\bf T}^{-1}
\end{array}\right],\quad {\bf P}={\bf D}-{\bf CA}^{-1}{\bf B}\,.
\eea
Substitution of the last expression into (\ref{DeformMetr}) gives
\bea
 ds^2=\frac{k}{4\pi}{\bf e}^T{\bf P}^{-T}
 \Big[{\bf H}{\bf H}^T-{\bf I}\Big]{\bf P}^{-1}{\bf e},\quad
 {e}_\alpha=-L_\alpha+[{\bf C}{\bf A}^{-1}]_{\alpha b}L_b\,.
 \eea 
 In particular, recalling that matrices $({\bf A},{\bf C})$ don't depend on the deformation, we recover the 
 expression (\ref{ds2ProdStr2}) for the deformation--independent ``bare'' frames. The information about the R--matrix is contained in 
\bea\label{Nov5b}
{\bf Q}\equiv {\bf P}^{-T}
 \Big[{\bf H}{\bf H}^T-{\bf I}\Big]{\bf P}^{-1},
\eea 
and in \cite{ChLGL} it was demonstrated every coset admits a ``canonical'' gauge, where matrix $Q$ has only constant entries. The goal of this appendix is to find the explicit form of such constant matrix for the $SO(N+1)/SO(N)$ cosets. First we observe that 
\bea\label{Nov5a}
{\bf P}={\bf H}^T-{\bf J}^T,
\eea
where matrix ${\bf J}$ does not depend on the deformation:
\bea\label{JdefApp}
J_{\alpha\beta}&=&D_{\alpha\beta}-D_{\alpha b}(D_{ab}-\delta_{ab})^{-1}D_{a\beta}
\eea
Substitution of (\ref{Nov5a}) into (\ref{Nov5b}) gives
\bea\label{Nov5c}
{\bf Q}={\bf I}+{\bf J}^T[{\bf H}-{\bf J}]^{-T}+[{\bf H}-{\bf J}]^{-1}{\bf J}+
{\bf P}^{-T}
 \Big[{\bf J}{\bf J}^T-{\bf I}\Big]{\bf P}^{-1}\,.
\eea
Matrix ${\bf J}$ for a given coset depends on the choice of the gauge, and in the remaining part of this appendix we will demonstrate that in the specific parameterization (\ref{gElemApp}) of $SO(N+1)/SO(N)$, matrix $J$ has a very simple form (\ref{JidentityApp}). Then ${\bf J}{\bf J}^T={\bf I}$, and the expression (\ref{Nov5c}) reduces to (\ref{QalBetAsJH})\footnote{One can show that the 
relation ${\bf J}{\bf J}^T={\bf I}$ holds in all gauges, but we will not discuss this further.}.

To compute the deformed dilaton (\ref{dilaton}), one should follow the procedure introduced for the standard lambda--deformation in \cite{DSTLmbAdS5} and rewrite the matrix (\ref{MatrM}) as 
\bea
M=
\left[\begin{array}{cc}
{\bf I}&{\bf C}\\
{\bf 0}&{\bf A}
\end{array}\right]
\left[\begin{array}{cc}
{\bf P}&{\bf 0}\\
{\bf A}^{-1}{\bf B}&{\bf I}
\end{array}\right]\,.\nonumber
\eea
This leads to the expression for the dilaton in terms of its undeformed value $e^{-2\Phi_0}$:
\bea\label{HJdilat}
e^{-2\Phi}=\mbox{det}M=[\mbox{det}{\bf A}][\mbox{det}{\bf P}]=
e^{-2\Phi_0}[\mbox{det}({\bf H}-{\bf J})]
\eea
Matrix ${\bf A}$ does not depend on the deformation, and the relation (\ref{JidentityApp}) implies that 
${\bf P}$ is a constant matrix. Thus the deformation multiplies the dilaton by an irrelevant constant. It is clear that the identity  (\ref{JidentityApp}) plays the central role in ensuring various analytical properties of the $\la$YB deformation, and the remaining part of this appendix is dedicated to proving the identity  (\ref{JidentityApp}) .

\bigskip

Let us now prove a very important property of the matrix $J$ defined by (\ref{JdefApp})\footnote{The anti--hermitian generators $T_M$ are assumed to be normalized by $\mbox{Tr}[T_MT_N]=-\delta_{MN}$.}.
Specifically, we will demonstrate that if one parameterizes the $SO(N+1)/SO(N)$ coset in terms of Euler angles as
 \bea\label{gElemApp}
 g=\underbrace{e^{\theta_N T_{N,N+1}}\dots e^{\theta_2 T_{2,3}}}_{h_L} \underbrace{e^{\theta_1 T_{1,2}}}_t\underbrace{e^{\theta_2 T_{2,3}}\dots e^{\theta_N T_{N,N+1}}}_{h_R}\,,
 \eea 
 then matrix $J$ takes a very simple form (\ref{Jidentity})
 \bea\label{JidentityApp}
J_{\alpha\beta}=(-1)^{\alpha+1}\delta_{\alpha\beta}.
\eea
This identity was discovered and verified in \cite{DSTLmbAdS5} for $N=3,4,5$, and here we present a recursive proof for all values of $N$.

\bigskip

We begin with analyzing the special case of $\theta_2=\dots=\theta_N$, and use ${\tilde D}$ to denote the corresponding matrix $D$. Using  the normalized generators of the coset and the subgroup,
\bea\label{BasOneOne}
T_\alpha=\frac{i}{\sqrt{2}}\{T_{1,2},\dots,T_{1,N+1}\}\,,\quad 
T_a=\frac{i}{\sqrt{2}}\{T_{2,3},\dots,T_{2,N+1};T_{3,4},\dots\}\,.
\eea 
as well as relations
\bea
e^{\theta_1 T_{12}}T_{1,p}e^{-\theta_1 T_{12}}=c_{\theta_1}T_{1,p}-s_{\theta_1}T_{2,p},\
e^{\theta_1 T_{12}}T_{2,p}e^{-\theta_1 T_{12}}=c_{\theta_1}T_{2,p}+s_{\theta_1}T_{1,p},\
p>2
\eea
we find the matrix $D$ in the basis (\ref{BasOneOne}):
\bea\label{defTildeJ}
{\tilde D}_{MN}=\left[\begin{array}{cc|cc}
1&0&0&0\\
0&c_{\theta_1}I_{N-1}&s_{\theta_1}I_{N-1}&0\\
\hline
0&-s_{\theta_1}I_{N-1}&c_{\theta_1}I_{N-1}&0\\
0&0&0&I_{\frac{N(N+1)}{2}-2N+1}
\end{array}\right]
\eea
To add other rotations, we define a set of matrices $M_k$ by 
\bea 
e^{-\theta_k T_{k,k+1}}T_M e^{\theta_k T_{k,k+1}}=(M_k)_{M,P}T_P,\quad k>1
\eea 
Action of such matrices on the generators $T_M$ produces a representation of the subgroup $F=SO(N)$, in particular, such action does not mix $T_\alpha$ with $T_a$. The actions of $h_L$ and $h_R$ defined in (\ref{gElemApp}) can be represented in terms of combinations of matrices $M_k$:
\bea 
h^{-1}_L T_M h_L =(M_N\cdots M_3 M_2)_{M,P}T_P,\quad 
h^{}_RT_N h^{-1}_R=(M_N^T\cdots M_3^TM_2^T)_{N,Q}T_Q\,,
\eea 
and matrix $D$ can be written as a product
\bea\label{DmnMrl}
 D_{MN}=\underbrace{(M_N\cdots M_3 M_2)}_{M_L}\tilde{D}\underbrace{(M_2M_3\cdots M_N)}_{M_R}
\eea 
Recalling that all matrices $M_k$ are block--diagonal (i.e., $M_{a\beta}=M_{\alpha b}=0$), we conclude that $M_L$ and $M_R$ have the same property, then the expression (\ref{JdefApp}) can be rewritten as
\bea\label{JrecurInterm}
J_{\gamma\delta}=M^L_{\gamma\alpha}\left[{\tilde D}_{\alpha\beta}-{\tilde D}_{\alpha b}\Big({\tilde D}_{ab}-(M_RM_L)_{ba}\Big)^{-1}{\tilde D}_{a\beta}\right]M^R_{\beta\delta}
\eea
Interestingly the term
 \be 
( M_R M_L)^T=(M_2M_3\cdots M_N^2\cdots M_3 M_2)^T=D_{MN}^{'}
 \ee 
is the $D$ matrix in a Euler parametrization of the coset $SO(N)/SO(N-1)$ with relabeled coordinates. Furthermore,
\bea 
D'_{ab}-{\tilde D}_{ab}=\left[\begin{array}{cc}
 	D_{\dot a\dot b}'-I_{\dot a\dot b} & D'_{\dot a\dot \beta}\\
 	D'_{\dot \alpha \dot b} &D'_{\dot \alpha\dot \beta}-\cos\theta_1\delta_{\dot \alpha\dot \beta}
 \end{array}\right]=\left[\begin{array}{cc}
 	D_{\dot a\dot b}'-I_{\dot a\dot b}&0\\
 		D'_{\dot \alpha \dot b}&I
  \end{array}\right]\left[\begin{array}{cc}
 	I&(D_{\dot a\dot b}'-I'_{\dot a\dot b})^{-1}D'_{\dot a\dot \beta}\\
 	0&J'_{\dot\alpha\dot\beta}-\cos\theta_1\delta'_{\dot\alpha\dot\beta}
 \end{array}\right]\nonumber
 \eea 
Here we separated the $SO(N)$ index $a$ into the $SO(N-1)$ subgroup and the coset $SO(N)/SO(N-1)$: $a=\{\dot a,\dot \alpha\}$. Substitution of the last relation into (\ref{JrecurInterm}) gives an expression for $J$ in terms of $J'$:
\bea\label{JinductRel}
J_{\gamma\delta}=M^L_{\gamma\alpha}\left[{\tilde D}_{\alpha\beta}-s_{\theta_1}^2(J'_{\dot\alpha\dot\beta}-c_{\theta_1}\delta'_{\dot\alpha\dot\beta})^{-1}\right]M^R_{\beta\delta}\equiv
M^L_{\gamma\alpha}{\hat J}_{\alpha\beta}M^R_{\beta\delta}\,.
\eea
We used the relations
\bea
{\tilde D}_{\alpha b}|_{b={\dot b}}=0,\quad {\tilde D}_{\alpha b}|_{b={\dot \beta}}=s_{\theta_1} \delta_{\alpha\dot\beta}\nonumber
\eea
that follow from (\ref{defTildeJ}). 

\bigskip

The recurrence relation (\ref{JinductRel}) is one of the main results of this appendix, and it leads to an inductive prove of the property (\ref{JidentityApp}):
\begin{enumerate}
\item{For $N=2$ and $\theta_2=0$, the property (\ref{JidentityApp}) follows from the expression (\ref{defTildeJ})\footnote{Recall that the matrix ${\hat J}_{\alpha\beta}$ is defined by (\ref{JinductRel}), and for $N=2$ it is obtained by setting $\theta_2=0$ in ${J}_{\alpha\beta}$.}:
\bea
{\hat J}_{\alpha\beta}=
\left[\begin{array}{cc}
1&0\\
0&c_{\theta_1}
\end{array}\right]-s_{\theta_1}^2\left[\begin{array}{cc}
1&0\\
0&\frac{1}{1-c_{\theta_1}}
\end{array}\right]=\left[\begin{array}{cc}
1&0\\
0&-1
\end{array}\right]
\eea
Transition from ${\hat J}_{\alpha\beta}$ to ${J}_{\alpha\beta}$ can be accomplished by performing explicit calculations with $\theta_2\ne 0$ or by implementing step 3 described below. This proves the relation (\ref{JidentityApp})  for $N=2$.
}
\item{Assuming that the property (\ref{JidentityApp}) holds for $SO(n+1)/SO(n)$ for $n<N$, we evaluate ${\hat J}_{\alpha\beta}$ for $SO(N+1)/SO(N)$ using (\ref{JinductRel}) and (\ref{defTildeJ}):
\bea
{\hat J}_{\alpha\beta}&=&\left[\begin{array}{cc}
1&0\\
0&c_{\theta_1}I_{N-1}
\end{array}\right]-s_{\theta_1}^2\left[\begin{array}{cc}
1&0\\
0&\frac{\delta_{\alpha'\beta'}}{(-1)^{\alpha'+1}-\cos\theta_1}
\end{array}\right]\nonumber\\
&=&\left[\begin{array}{cc}
1&0\\
0&-(-1)^{\alpha'+1}\delta_{\alpha'\beta'}
\end{array}\right]=(-1)^{\alpha+1}\delta_{\alpha\beta}\,.
\eea
This implies that matrix ${\hat J}_{\alpha\beta}$ satisfies the identity (\ref{JidentityApp}).
}
\item{To extend the result (\ref{JidentityApp}) from ${\hat J}_{\alpha\beta}$ to ${J}_{\alpha\beta}$, we observe that the matrices $M_k$ comprising $(M_L,M_R)$ (see (\ref{DmnMrl})) have the form
\bea
(M_k)_{\alpha\beta}=\left[\begin{array}{ccc}
I&0&0\\
	0&\begin{pmatrix}
	\cos\theta_k&\sin\theta_k\\
	-\sin\theta_k&\cos\theta_k
	\end{pmatrix}_{k,k+i}&0\\
	0&0&I
\end{array}\right]
\eea
Then $M_k{\hat J}M_k={\hat J}$, and repeated application of this relation leads to the conclusion the $J=M_L {\hat J}M_R={\hat J}$ for $SO(N+1)/SO(N)$, which completes the inductive proof of  the identity (\ref{JidentityApp}).
}
\end{enumerate}

\noindent
To summarize, in this appendix we have outlined the derivation of the expressions (\ref{ds2ProdStr})--(\ref{QalBetAsJH}) for the deformed metric and proved the identity (\ref{JidentityApp}) for $SO(N+1)/SO(N)$, which is used in section \ref{SectMainDeform} to construct the $\la$YB deformation of the metric on a sphere.

\section{$B$--field for the deformed geometries}
\label{AppBzero}

In this appendix we will demonstrate that the integrable deformations constructed in section \ref{SectMainDeform} have vanishing Kalb--Ramond field.  Our proof will follow the logic developed in \cite{GrigTseytl}.

We begin with we recalling the action (\ref{GenLambdaAction})
\bea\label{GenLambdaActionV2}
S_{k,\lambda}(g)=S_{WZW,k}(g)+\frac{k}{2\pi}\int d^2\sigma L_+^a[\hat{\lambda}^{-1}-D]^{-1}R_-^b,
\eea
and observing that the first term given by (\ref{gWZW}) does not depend on the deformation. Since the undeformed WZW model on AdS$_p\times$S$^q$ had $B=0$, the Kalb--Ramond field can come only from the second term in (\ref{GenLambdaActionV2}), so we should focus on the antisymmetric part of this term:
\bea\label{tempBeqn1}
\mathcal{L}_B=L_+({\hat\lambda}^{-1}-D)^{-1}D L_--L_-({\hat\lambda}^{-1}-D)^{-1}D L_+
\eea
To prove that $\mathcal{L}_B=0$, we use the automorphism of the Lie algebra introduced in \cite{GrigTseytl}:
\be\label{AutoMap}
T_{MN}\rightarrow \tilde{T}_{MN}=T_{MN}(-1)^{M+N}
\ee
It is clear that this map preserves the commutation relations (\ref{SOalgebraMain})\footnote{In this appendix we are focusing on $SO(N+1)$ part of the sigma model, and the AdS part can be treated in the same way.}. While the transformation (\ref{AutoMap}) leaves the {\it family} of integrable deformations invariant, to argue that $\mathcal{L}_B=0$ we need to show that the sigma model with a specific 
${\hat\lambda}^{-1}$ is mapped to itself rather than to a solution with $[{\hat\lambda}^{-1}]'$. To show this we recall the action (\ref{RmatrCanon}) of the canonical R--matrix and observe that under the automorphism (\ref{AutoMap}) leads to a map
\bea\label{RtildeMapAuto}
\r\rightarrow {\tilde\r}^T,\quad {\tilde R}_{sq}=R_{qs},
\eea
where ${\tilde\r}$ is still given by (\ref{RmatrCanon}), but with a different Cartan block. The canonical R--matrix appears in the deformation only through combination (\ref{RinvInterm}),
\bea
\left[(1-\eta\RR)^{-1}\right]_{\alpha\beta}=
\left[
\begin{array}{cc}
\frac{1}{1+\gamma\eta^2}&0\\
0&\frac{1}{1+\eta^2}I_{N-1}
\end{array}
\right],\qquad \gamma=V^T(1-\eta A)^{-1} V\,.
\eea
which remains invariant under the map (\ref{RtildeMapAuto}) since ${\tilde V}=V$, ${\tilde A}=A$. Thus we have demonstrated that the action (\ref{GenLambdaActionV2}) is invariant under the automorphism (\ref{AutoMap}).

Since each term in (\ref{tempBeqn1}) is invariant under (\ref{AutoMap}), application of the automorphism only to the first term does not change the Lagrangian. In the Euler angle parameterization (\ref{gElemApp}), various ingredients defined in (\ref{lambdaMtrDef}) transform as 
\bea 
g\rightarrow \tilde{g}=g^{-1},\quad L_M\rightarrow
\tilde{L}_M=-(-1)^{M}R_M,\quad
D_{MN}\rightarrow \tilde{D}_{MN}=(-1)^{M+N}D^T_{MN}
\eea
Introducing a convenient matrix $K$:
\bea
K_{MN}=(-1)^M \delta_{MN},
\eea
we arrive at a transformation\footnote{We used the identity $DD^T=I$.}
\bea
L_+({\hat\lambda}^{-1}-D)^{-1}D L_-\rightarrow R_+[K{\hat{\tilde\la}}\,^{-1}K-D^T]^{-1}D^T R_-=
L_-[K{\hat{\tilde\la}}\,^{-T}K-D]^{-1}DL_+
\eea
Since each term in (\ref{tempBeqn1}) must be invariant under the automorphism, the Lagrangian can be rewritten as
\bea\label{tempBeqn2}
\mathcal{L}_B=L_-(K{\hat{\tilde\la}}\,^{-T}K-D)^{-1}D L_+-L_-({\hat\lambda}^{-1}-D)^{-1}D L_+\,.
\eea
As we demonstrated in section \ref{SectMainDeform}, the canonical solution of the Yang--Baxter equation leads to a diagonal matrix $\hat\la={\hat{\tilde\la}}$, then $K{\hat\lambda}^{-1}K={\hat\lambda}^{-1}$ and the $B$--field described by (\ref{tempBeqn2}) vanishes. 

To extend this result to the general R--matrix, we recall the expression (\ref{Eqn420}):
\bea\label{BfldEqn7}
[{\hat\lambda}^{-1}]_{\alpha\beta}=[f_1 {\hat\lambda}^{-1}_{can} f_1^{-1}]_{\alpha\beta},
\eea
where ${\hat\lambda}^{-1}_{can}$ is the deformation parameter for the canonical R--matrix. Rotation $f_1$ and its image under (\ref{AutoMap}) transform the basis in the same way:
\bea
f_1 T_{m}f_1^T=h_{mn}T_n,\quad \tilde{f}_1\tilde{T}_m\tilde{f}_1^T=h_{mn}\tilde{T}_n,
\eea 
then we find 
\be 
\tilde{h}_{mn}=\text{Tr}[\tilde{T}_m\tilde{f}_1\tilde{T}_n\tilde{f}_1^T ]=[K hK]_{mn}.
\ee 
Application of this relation to (\ref{BfldEqn7}) gives
\be 
[{\hat\lambda}^{-1}]_{\alpha\beta}\rightarrow 
K [f_1 {\hat\lambda}^{-1}_{can} f_1^{-1}]_{\alpha\beta} K=K[{\hat\lambda}^{-1}]_{\alpha\beta}K
\ee 
Substitution of this relation into (\ref{tempBeqn2}) leads to $\mathcal{L}_B=0$, concluding the proof that the Kalb--Ramond field indeed vanishes for the generalized $\la$--deformation of $S^N$.


\begin{thebibliography}{30}

\bibitem{AdS5int}
 J.~A.~Minahan and K.~Zarembo,
  ``The Bethe ansatz for N=4 superYang-Mills,''
  JHEP {\bf 0303}, 013 (2003), hep-th/0212208;\\
I.~Bena, J.~Polchinski and R.~Roiban,
  ``Hidden symmetries of the AdS(5) x S**5 superstring,''
  Phys.\ Rev.\ D {\bf 69}, 046002 (2004),
  hep-th/0305116.
  %
\bibitem{AdS3int}
A.~Babichenko, B.~Stefanski, Jr. and K.~Zarembo,
  ``Integrability and the AdS(3)/CFT(2) correspondence,''
  JHEP {\bf 1003}, 058 (2010), arXiv:0912.1723;\\
%
A.~Cagnazzo and K.~Zarembo,
  ``B-field in AdS(3)/CFT(2) Correspondence and Integrability,''
  JHEP {\bf 1211}, 133 (2012), arXiv:1209.4049;\\
  %
  B.~Hoare and A.~A.~Tseytlin,
  ``On string theory on AdS(3) x S(3) x T(4) with mixed 3-form flux: tree-level S-matrix,''
  Nucl.\ Phys.\ B {\bf 873}, 682 (2013),  arXiv:1303.1037;\\
  %
  A.~Sfondrini,
  ``Towards integrability for ${\rm Ad}{{{\rm S}}_{{\bf 3}}}/{\rm CF}{{{\rm T}}_{{\bf 2}}}$,''
  J.\ Phys.\ A {\bf 48}, 023001 (2015), arXiv:1406.2971,\\
  %
  R.~Borsato, O.~Ohlsson Sax, A.~Sfondrini, B.~Stefanski and A.~Torrielli,
  ``The all-loop integrable spin-chain for strings on AdS$_3 \times S^3 \times T^4$: the massive sector,''
  JHEP {\bf 1308}, 043 (2013), arXiv:1303.5995;\\
B.~Hoare and A.~A.~Tseytlin,
  ``Massive S-matrix of AdS3 x S3 x T4 superstring theory with mixed 3-form flux,''
  Nucl.\ Phys.\ B {\bf 873}, 395 (2013), arXiv:1304.4099;\\
R.~Borsato, O.~Ohlsson Sax, A.~Sfondrini, B.~Stefanski, Jr. and A.~Torrielli,
  ``Dressing phases of AdS3/CFT2,''
  Phys.\ Rev.\ D {\bf 88}, 066004 (2013), arXiv:1306.2512;\\
 R.~Borsato, O.~Ohlsson Sax, A.~Sfondrini and B.~Stefanski,
  ``Towards the All-Loop Worldsheet S Matrix for $AdS_3\times S^3\times T^4$,''
  Phys.\ Rev.\ Lett.\  {\bf 113}, no. 13, 131601 (2014). arXiv:1403.4543;
 ``The complete AdS$_{3} \times$ S$^3 \times$ T$^4$ worldsheet S matrix,''
  JHEP {\bf 1410}, 66 (2014), arXiv:1406.0453;\\
T.~Lloyd, O.~Ohlsson Sax, A.~Sfondrini and B.~Stefanski, Jr.,
  ``The complete worldsheet S matrix of superstrings on AdS3 x S3 x T4 with mixed three-form flux,''
  Nucl.\ Phys.\ B {\bf 891}, 570 (2015), arXiv:1410.0866;\\
R.~Borsato, O.~Ohlsson Sax, A.~Sfondrini, B.~Stefanski, Jr. and A.~Torrielli,
  arXiv:1607.00914 [hep-th].

\bibitem{AdS2int}
D.~Sorokin, A.~Tseytlin, L.~Wulff and K.~Zarembo,
  ``Superstrings in AdS$_2\times$S$^2\times$T$^6$,''
  J.\ Phys.\ A {\bf 44}, 275401 (2011), arXiv:1104.1793;\\
%
L.~Wulff,
  ``Superisometries and integrability of superstrings,''
  JHEP {\bf 1405}, 115 (2014),
  arXiv:1402.3122;
  ``On integrability of strings on symmetric spaces,''
  JHEP {\bf 1509}, 115 (2015),
  arXiv:1505.03525.

 \bibitem{beta}
 R.~Roiban,
  ``On spin chains and field theories,''
  JHEP {\bf 0409}, 023 (2004),
  hep-th/0312218;\\
O.~Lunin and J.~M.~Maldacena,
  ``Deforming field theories with U(1) x U(1) global symmetry and their gravity duals,''
  JHEP {\bf 0505}, 033 (2005), hep-th/0502086;\\
S.~A.~Frolov, R.~Roiban and A.~A.~Tseytlin,
  ``Gauge-string duality for superconformal deformations of N=4 super Yang-Mills theory,''
  JHEP {\bf 0507}, 045 (2005), hep-th/0503192;\\
S.~Frolov,
  ``Lax pair for strings in Lunin-Maldacena background,''
  JHEP {\bf 0505}, 069 (2005), hep-th/0503201;\\
N.~Beisert and R.~Roiban,
  ``Beauty and the twist: The Bethe ansatz for twisted N=4 SYM,''
  JHEP {\bf 0508}, 039 (2005),  hep-th/0505187;\\
%
 S.~A.~Frolov, R.~Roiban and A.~A.~Tseytlin,
  ``Gauge-string duality for (non)supersymmetric deformations of N=4 super Yang-Mills theory,''
  Nucl.\ Phys.\ B {\bf 731}, 1 (2005), hep-th/0507021.


\bibitem{DelducEta}
 F.~Delduc, M.~Magro and B.~Vicedo,
  ``An integrable deformation of the AdS$_5$ x S$^5$ superstring action,''
  Phys.\ Rev.\ Lett.\  {\bf 112}, no. 5, 051601 (2014), arXiv:1309.5850;\\
  ``Derivation of the action and symmetries of the $q$-deformed $AdS_{5} \times S^{5}$ superstring,''
  JHEP {\bf 1410}, 132 (2014), arXiv:1406.6286.

  \bibitem{eta}
G.~Arutyunov, R.~Borsato and S.~Frolov,
  ``S-matrix for strings on $\eta$-deformed AdS5 x S5,''
  JHEP {\bf 1404}, 002 (2014)
  arXiv:1312.3542;\\
  %
B.~Hoare, R.~Roiban and A.~A.~Tseytlin,
 ``On deformations of $AdS_n$ x $S^n$ supercosets,''
  JHEP {\bf 1406}, 002 (2014)
  arXiv:1403.5517;\\
  %
  O.~Lunin, R.~Roiban and A.~A.~Tseytlin,
  ``Supergravity backgrounds for deformations of AdS$_{n} \times S^n$ supercoset string models,''
  Nucl.\ Phys.\ B {\bf 891}, 106 (2015), arXiv:1411.1066;
  \\
  %
 B.~Hoare,
  ``Towards a two-parameter q-deformation of AdS$_3 \times S^3 \times M^4$ superstrings,''
  Nucl.\ Phys.\ B {\bf 891}, 259 (2015), arXiv:1411.1266;\\
  %
 S.~J.~van Tongeren,
  ``On classical Yang-Baxter based deformations of the AdS$_{5}$ × S$^{5}$ superstring,''
  JHEP {\bf 1506}, 048 (2015), arXiv:1504.05516;\\
 G.~Arutyunov, R.~Borsato and S.~Frolov,   
 ``Puzzles of $\eta$-deformed AdS$_5 \times$ S$^5$,''
  JHEP {\bf 1512}, 049 (2015)
  arXiv:1507.04239;
  %
  \\
G.~Arutyunov, S.~Frolov, B.~Hoare, R.~Roiban and A.~A.~Tseytlin,
  ``Scale invariance of the $\eta$-deformed $AdS_5\times S^5$ superstring, T-duality and modified type II equations,''
  Nucl.\ Phys.\ B {\bf 903}, 262 (2016),
  arXiv:1511.05795;\\
  %
  %
  %
L.~Wulff and A.~A.~Tseytlin,
  ``Kappa-symmetry of superstring sigma model and generalized 10d supergravity equations,''
  arXiv:1605.04884;\\
R.~Borsato and L.~Wulff,
  ``Target space supergeometry of $\eta$ and $\lambda$-deformed strings,''
  arXiv:1608.03570 [hep-th].
 

\bibitem{SftsGr} 
  K.~Sfetsos,
  ``Integrable interpolations: From exact CFTs to non-Abelian T-duals,''
  Nucl.\ Phys.\ B {\bf 880}, 225 (2014)
  arXiv:1312.4560.
  
\bibitem{MoreLmb} 
 T.~J.~Hollowood, J.~L.~Miramontes and D.~M.~Schmidtt,
  ``Integrable Deformations of Strings on Symmetric Spaces,''
  JHEP {\bf 1411}, 009 (2014), 
  arXiv:1407.2840;\\
  %
  T.~J.~Hollowood, J.~L.~Miramontes and D.~M.~Schmidtt,
  ``An Integrable Deformation of the $AdS_5 \times S^5$ Superstring,''
  J.\ Phys.\ A {\bf 47}, no. 49, 495402 (2014)
  arXiv:1409.1538.
  K.~Sfetsos and D.~C.~Thompson,
  ``Spacetimes for $\lambda$-deformations,''
  JHEP {\bf 1412}, 164 (2014)
  arXiv:1410.1886;\\
%
C.~Appadu and T.~J.~Hollowood,
  ``Beta function of k deformed AdS$_{5}$ x S$^{5}$ string theory,''
  JHEP {\bf 1511} (2015) 095, arXiv:1507.05420;\\
%
  B.~Hoare and A.~A.~Tseytlin,
  ``On integrable deformations of superstring sigma models related to $AdS_n \times S^n$ supercosets,''
  Nucl.\ Phys.\ B {\bf 897}, 448 (2015)
  arXiv:1504.07213;\\
  R.~Borsato, A.~A.~Tseytlin and L.~Wulff,
  ``Supergravity background of $\lambda$-deformed model for AdS$_2 \times$  S$^2$ supercoset,''
  Nucl.\ Phys.\ B {\bf 905}, 264 (2016),
  arXiv:1601.08192;\\
  Y.~Chervonyi and O.~Lunin,
  ``Supergravity background of the $\lambda$-deformed AdS$_3 \times$ S$^3$ supercoset,''
  Nucl.\ Phys.\ B {\bf 910}, 685 (2016), arXiv:1606.00394 [hep-th];\\
 R.~Borsato and L.~Wulff,
  ``Target space supergeometry of $\eta$ and $\lambda$-deformed strings,''
  JHEP {\bf 1610}, 045 (2016), arXiv:1608.03570 [hep-th].
  
\bibitem{DSTLmbAdS5}
  S.~Demulder, K.~Sfetsos and D.~C.~Thompson,
  ``Integrable $\lambda$-deformations: Squashing Coset CFTs and $AdS_5\times S^5$,''
  JHEP {\bf 1507}, 019 (2015), arXiv:1504.02781.
 
\bibitem{GenLambda}
  K.~Sfetsos, K.~Siampos and D.~C.~Thompson,
  ``Generalised integrable $\lambda$-- and $\eta$-deformations and their relation,''
  Nucl.\ Phys.\ B {\bf 899}, 489 (2015), arXiv:1506.05784 [hep-th].
  
\bibitem{Cherednyk} 
  I.~V.~Cherednik,
  ``Relativistically Invariant Quasiclassical Limits of Integrable Two-dimensional Quantum Models,''
  Theor.\ Math.\ Phys.\  {\bf 47}, 422 (1981).
  
 \bibitem{BosonicYangBaxter}
C.~Klimcik,
  ``Yang-Baxter sigma models and dS/AdS T duality,''
  JHEP {\bf 0212}, 051 (2002), hep-th/0210095; \\
  %
C.~Klimcik,
  ``On integrability of the Yang-Baxter sigma-model,''
  J.\ Math.\ Phys.\  {\bf 50}, 043508 (2009), arXiv:0802.3518;\\
  %
C.~Klimcik,
 ``Integrability of the bi-Yang-Baxter sigma-model,''
  Lett.\ Math.\ Phys.\  {\bf 104}, 1095 (2014), arXiv:1402.2105;\\
C.~Klimcik,
  ``Poisson--Lie T-duals of the bi-Yang--Baxter models,''
  Phys.\ Lett.\ B {\bf 760}, 345 (2016),
  arXiv:1606.03016 [hep-th].

\bibitem{Qdeform}  F.~Delduc, M.~Magro and B.~Vicedo,
  ``On classical $q$-deformations of integrable sigma-models,''
  JHEP {\bf 1311}, 192 (2013), arXiv:1308.3581.


\bibitem{Yoshida}
I.~Kawaguchi, T.~Matsumoto and K.~Yoshida,
  ``Jordanian deformations of the $AdS_5 x S^5$ superstring,''
  JHEP {\bf 1404}, 153 (2014), arXiv:1401.4855;\\
T.~Matsumoto and K.~Yoshida,
  ``Yang--Baxter sigma models based on the CYBE,''
  Nucl.\ Phys.\ B {\bf 893}, 287 (2015),  arXiv:1501.03665;\\
T.~Kameyama, H.~Kyono, J.~i.~Sakamoto and K.~Yoshida,
  ``Lax pairs on Yang--Baxter deformed backgrounds,''
  JHEP {\bf 1511}, 043 (2015), arXiv:1509.00173;\\
B.~Hoare and S.~J.~van Tongeren,
  ``On jordanian deformations of AdS5 and supergravity,''
  arXiv:1605.03554;\\
H.~Kyono and K.~Yoshida,
  ``Supercoset construction of Yang-Baxter deformed AdS$_5\times$S$^5$ backgrounds,''
  arXiv:1605.02519;\\
D.~Orlando, S.~Reffert, J.~i.~Sakamoto and K.~Yoshida,
  ``Generalized type IIB supergravity equations and non-Abelian classical r-matrices,''
  arXiv:1607.00795.
  
 \bibitem{ChLGL} 
  Y.~Chervonyi and O.~Lunin,
  ``Generalized $\lambda$-deformations of AdS$_p \times$ S$^p$,''
  Nucl.\ Phys.\ B {\bf 913}, 912 (2016),
  arXiv:1608.06641 [hep-th].
 
\bibitem{BelavDrinf}
Belavin, A. A., Drinfel'd, V. G., ``Solutions of the classical Yang-Baxter equation for simple Lie
algebras'', Funct. Anal. Appl. 16, 159 (1982).

  
\bibitem{CYB}
P.~P.~Kulish, N.~Y.~Reshetikhin and E.~K.~Sklyanin,
  ``Yang-Baxter Equation and Representation Theory. 1.,''
  Lett.\ Math.\ Phys.\  {\bf 5}, 393 (1981);
 \\
  E.~K.~Sklyanin,
  ``Some algebraic structures connected with the Yang-Baxter equation,''
  Funct.\ Anal.\ Appl.\  {\bf 16}, 263 (1982);
  \\
M.~A.~Semenov-Tian-Shansky,
  ``What is a classical r-matrix?,''
  Funct.\ Anal.\ Appl.\  {\bf 17}, 259 (1983);
  \\
Drinfel'd, V. G., ``Hamiltonian structures on Lie groups, Lie bi-algebras and the geometric meaning of the classical Yang-Baxter equations,'' Sov. Math. Dokl. 27, 68 (1983);\\
%
  M.~A.~Semenov-Tian-Shansky,
  ``Dressing transformations and Poisson group actions,''
  Publ.\ Res.\ Inst.\ Math.\ Sci.\ Kyoto {\bf 21}, 1237 (1985).
  
\bibitem{WZW} 
  E.~Witten,
  ``Nonabelian Bosonization in Two-Dimensions,''
  Commun.\ Math.\ Phys.\  {\bf 92}, 455 (1984).

\bibitem{PCM}
  A.~M.~Polyakov,
  ``Interaction of Goldstone Particles in Two-Dimensions. Applications to Ferromagnets and Massive Yang-Mills Fields,''
  Phys.\ Lett.\ B {\bf 59}, 79 (1975).
  
\bibitem{xxx} 
G.~Georgiou, K.~Sfetsos and K.~Siampos,
  ``$\lambda$-Deformations of left–right asymmetric CFTs,''
  Nucl.\ Phys.\ B {\bf 914}, 623 (2017), arXiv:1610.05314 [hep-th];\\
G.~Georgiou and K.~Sfetsos,
  ``A new class of integrable deformations of CFTs,''
  JHEP {\bf 1703}, 083 (2017), arXiv:1612.05012 [hep-th];\\
J.~i.~Sakamoto and K.~Yoshida,
  ``Yang-Baxter deformations of $W_{2,4}\times T^{1,1}$ and the associated T-dual models,''
  Nucl.\ Phys.\ B {\bf 921}, 805 (2017)
  arXiv:1612.08615 [hep-th];\\
J.~i.~Sakamoto, Y.~Sakatani and K.~Yoshida,
  ``Homogeneous Yang-Baxter deformations as generalized diffeomorphisms,''
  J.\ Phys.\ A {\bf 50}, 415401 (2017), arXiv:1705.07116;\\
 G.~Georgiou and K.~Sfetsos,
  ``Integrable flows between exact CFTs,''
  arXiv:1707.05149;\\
 F.~Delduc, B.~Hoare, T.~Kameyama and M.~Magro,
  ``Combining the bi-Yang-Baxter deformation, the Wess-Zumino term and TsT transformations in one integrable sigma-model,''
  arXiv:1707.08371 [hep-th];\\
B.~Hoare and F.~K.~Seibold,
  ``Poisson-Lie duals of the eta deformed symmetric space sigma model,''
  arXiv:1709.01448 [hep-th];\\
J.~J.~Fernandez-Melgarejo, J.~i.~Sakamoto, Y.~Sakatani and K.~Yoshida,
  ``$T$-folds from Yang-Baxter deformations,''
  arXiv:1710.06849 [hep-th];\\
 S.~Demulder, S.~Driezen, A.~Sevrin and D.~C.~Thompson,
  ``Classical and Quantum Aspects of Yang-Baxter Wess-Zumino Models,''
  arXiv:1711.00084 [hep-th].

\bibitem{SWmap}
T.~Araujo, I.~Bakhmatov, E.~O.~Colgain, J.~Sakamoto, M.~M.~Sheikh-Jabbari and K.~Yoshida,
  ``Yang-Baxter $\sigma$-models, conformal twists, and noncommutative Yang-Mills theory,''
  Phys.\ Rev.\ D {\bf 95}, no. 10, 105006 (2017), arXiv:1702.02861 [hep-th];\\
 T.~Araujo, I.~Bakhmatov, E.~O.~Colgain, J.~i.~Sakamoto, M.~M.~Sheikh-Jabbari and K.~Yoshida,
  ``Conformal Twists, Yang-Baxter $\sigma$-models \& Holographic Noncommutativity,''
  arXiv:1705.02063 [hep-th];\\
T.~Araujo, E.~O Colgain, J.~Sakamoto, M.~M.~Sheikh-Jabbari and K.~Yoshida,
  ``$I$ in generalized supergravity,''
  Eur.\ Phys.\ J.\ C {\bf 77} 739 (2017),  arXiv:1708.03163 [hep-th];\\
I.~Bakhmatov, O.~Kelekci, E.~O.~Colgain and M.~M.~Sheikh-Jabbari,
  ``Classical Yang-Baxter Equation from Supergravity,''
  arXiv:1710.06784 [hep-th].
  
\bibitem{Guide}
 V.~Chari and A.~Pressley,
  ``A guide to quantum groups,''
  Cambridge, UK: Univ. Pr. (1994) 651 p
  
 \bibitem{Bars}
I.~Bars and K.~Sfetsos,
  ``A Superstring theory in four curved space-time dimensions,''
  Phys.\ Lett.\ B {\bf 277}, 269 (1992),
  hep-th/9111040.

\bibitem{GrigTseytl}
M.~Grigoriev and A.~A.~Tseytlin,
  ``Pohlmeyer reduction of AdS(5) x S$^5$ superstring sigma model,''
  Nucl.\ Phys.\ B {\bf 800}, 450 (2008), arXiv:0711.0155 [hep-th].
  
\end{thebibliography}
 \end{document}